\documentclass[aps,prl,twocolumn,psfig,groupedaddress]{revtex4}
\def\bc{\begin{center}}
\def\ec{\end{center}}
\def\be{\begin{equation}}
\def\ee{\end{equation}}
\usepackage{longtable}
\usepackage{psfig}

\begin{document}

%\preprint{}
\title{Composite fermions in the neighborhood of $\nu=1/3$}
\author{Jainendra K. Jain, Chia-Chen Chang, Gun Sang Jeon, 
and Michael R. Peterson}
%\author{J.K. Jain}
%\thanks{}
\affiliation{Physics Department, 104 Davey Laboratory, The Pennsylvania State University,
University Park, Pennsylvania 16802}

\date{May 14, 2003}

\begin{abstract}
We present extensive comparisons of the composite fermion theory with exact 
results in the filling factor range $2/5>\nu>1/3$, which affirm that   
the composite fermion theory correctly describes the qualitative  
reorganization of the low energy Hilbert space of the strongly correlated 
electrons, and predicts eigenenergies with an accuracy of $\sim 0.1$ \%. 
These facts establish the basic validity of the composite 
fermion description in this filling factor region. 
\end{abstract}
\pacs{71.10.Pm,73.43.-f}
%\keywords{}


\maketitle

This article concerns the applicability of the composite fermion theory in 
the filling factor region $2/5 > \nu > 1/3$, which has come into focus because of 
the recent observation\cite{Pan,Smet} of fractional quantum Hall effect (FQHE)\cite{Tsui} 
at several fractions in this range, for example $\nu=4/11$.  
Given that the CF theory\cite{Jain,Heinonen,Pinczuk1} 
provides a unified and quantitatively accurate description 
of the FQHE liquid in a large range of lowest Landau level (LL) fillings, 
it might seem natural that it also 
captures the physics in the filling factor range $2/5 > \nu > 1/3$.   
This article will review earlier theoretical studies 
and present more extensive recent theoretical tests which confirm 
the validity of the CF description in this parameter region.

\section{Composite fermion basics}

\begin{figure}
\centerline{\psfig{figure=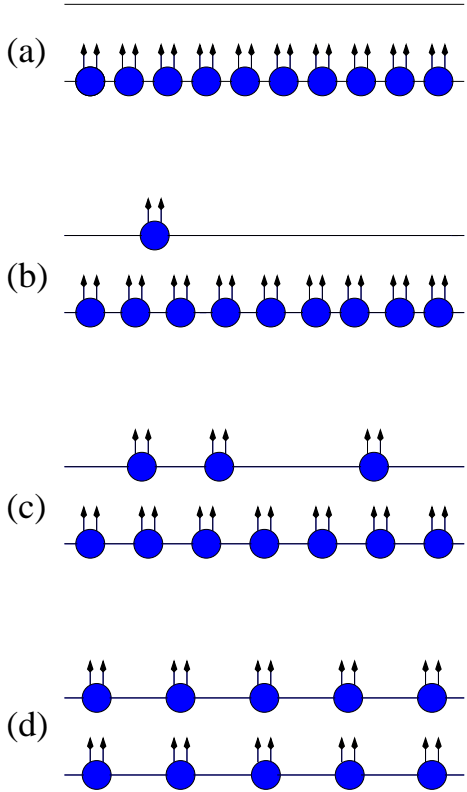,width=3.0in,angle=-90}}
\caption{The physical picture for the 
evolution of the many-body state as the filling factor varies between $\nu=1/3$ 
and $\nu=2/5$ (a and d, respectively).  Panel b shows a solitary composite fermion
in the second level, which is interpreted as a ``quasiparticle" of $\nu=1/3$.
Panel c shows an intermediate filling factor in the range $2/5>\nu>1/3$.}
\label{fig0}
\end{figure}

The central principle of the composite fermion theory is that 
interacting electrons at filling factor $\nu$ minimize their interaction 
energy by transforming into 
weakly interacting composite fermions.  The composite fermions experience
an effective magnetic field, and have a filling factor $\nu^*$,
where $\nu$ and $\nu^*$ are related by $\nu=\nu^*/(2p\nu^* \pm 1)$,
with the even integer $2p$ denoting the vorticity of the composite fermion.
The formation of composite fermions leads to a profound reorganization 
of the low energy Hilbert space of interacting electrons.
In particular, a gap opens up at $\nu^*=n$, where $n$ is an integer, which
immediately explains the FQHE of electrons at $\nu=n/(2pn\pm 1)$ as  
the integral quantum Hall effect (IQHE)\cite{Klitzing} of composite fermions.  
It is then natural to describe the  
region between $\frac{(n+1)}{2p(n+1)\pm 1} > \nu > \frac{n}{2pn\pm 1}$
in terms of a CF state for which
$n$ CF levels are fully occupied and the $(n+1)^{st}$ level is partially
occupied.  Fig.~(\ref{fig0}) displays the physical picture in the range 
$2/5 \geq \nu \geq 1/3$.

Briefly, our method for calculating the low energy spectrum is as follows.  
We will denote the total
number of composite fermions (which is the same as the total number of
electrons) by $N$, the CF filling factor by $\nu^*=n+\bar \nu$,
and the number of composite fermions in the topmost
partially occupied level by $\bar N$.  We take 
all basis states of {\em electrons} at filling $\nu^*=n+\bar \nu$ in which the 
$n$ lowest electronic LLs are fully occupied and the $(n+1)^{st}$ LL 
has filling $\bar \nu$.  (States involving  excitations of electrons across 
a LL are neglected at the lowest order approximation.)  From this
basis, we construct a basis for composite fermions at $\nu^*=n+\bar \nu$
following the standard procedure \cite{JK}.  The advantage of going 
to the CF description is that the dimension of the Fock space 
in the CF theory is exponentially small compared to the dimension of the 
Fock space of original electron problem.  Indeed, when  $\bar \nu=0$, 
i.e., $\nu^*=n$, there is only a single state in the low energy band, and nothing 
further remains to be done (except, of course, to evaluate its 
properties).  This state is the $\nu=n/(2pn+1)$ FQHE ground state.
When there are many basis states, we diagonalize the Coulomb Hamiltonian in the 
CF basis.
We will only give the results here, referring the reader to the literature for 
the construction of the wave functions and the diagonalization 
procedure \cite{JK,Mandal}.  It is stressed that all energies obtained in 
our calculations are exact variational upper bounds.
(A related approach for treating interacting composite fermions 
\cite{Quinn1,Quinn2,Lee} formulates the problem in terms of fermions with 
pair-wise interaction, which is deduced either from exact 
diagonalization or from CF theory by considering a system which has only two 
composite fermions in the second CF level.)

Our focus in this article will be on the filling factor range $2/5>\nu>1/3$.
It has been confirmed in the past that the 
states at $\nu=1/3$ and $2/5$ are well described as one and two 
filled level of composite fermions.  In this article, we ask if  
the composite fermion description continues to be valid at intermediate 
filling factors, where the composite fermions filling factor is $\nu^*=1+\bar \nu$.
The underlying physical picture is shown in Fig.~(\ref{fig0}): as the 
filling factor is increased from $\nu=1/3$ ($\nu^*=1$), the  
second level of composite fermions gets increasingly more densely populated,
eventually becoming completely full and producing incompressibility at 
$\nu=2/5$ ($\nu^*=2$).

The spherical geometry is used in all our calculations below, in which 
electrons move on the surface of a sphere, with the total flux through the 
surface of the sphere given by $2Q$ in units of the quantum of flux $\phi_0=hc/e$, 
where $Q$ is 
either an integer or a half integer. The Landau levels are angular momentum 
shells in this geometry, with the lowest LL corresponding 
to angular momentum $Q$, the second to $Q+1$, and so on.
The many body eigenstates are labeled by their total orbital angular momentum
quantum number $L$.  Interacting electrons at $Q$ map into composite fermions
at $Q^*=Q-p(N-1)$.  It will be assumed that the magnetic field is sufficiently 
large that the spin degree of freedom is frozen and only fully spin polarized
states are relevant. 

\section{Excitations at $\nu=1/3$}

The electron filling factor $\nu=1/3$ maps into filling factor $\nu^*=1$ of 
composite fermions: the ground state is interpreted as one filled 
Landau level of composite fermions (Fig.~\ref{fig0}a) and excited states are 
obtained by creating particle hole pairs of composite fermions.

The effective flux at $\nu^*=1$ is $2Q^*=N-1$.  For the lowest energy excitation,
one particle is promoted to the next level, thus creating a particle 
hole pair of composite fermions.  The particle and the hole have 
angular momenta $Q^*+1=(N+1)/2$ and $Q^*=(N-1)/2$, respectively, giving 
the total angular momentum values of $L=1,2, \cdots N$ for the pair.   
It turns out that the particle hole state 
with $L=1$ at $\nu^*=1$ does not produce any $L=1$ state for composite
fermions; the wave function is annihilated upon lowest LL projection \cite{Dev}.
Thus the CF theory predicts single multiplets at $L=2, \cdots N$, which 
is in agreement with exact results.  The energies (see $E_{CF}^{(1)}$ 
in Table I) are in good agreement with the exact energies, especially in 
view of the fact that the wave functions contain no adjustable parameters.

\begin{figure}
\centerline{\psfig{figure=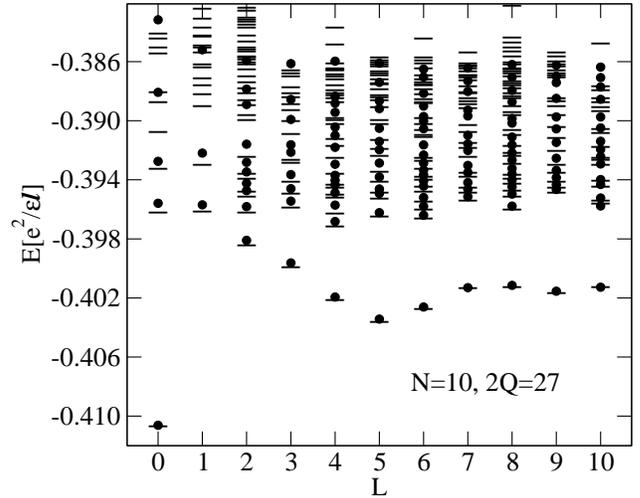,width=3.5in,angle=0}}
\vspace{-0.5cm}
\caption{The dots show the spectrum of low energy states predicted by the 
CF theory for $N=10$ electrons at flux $2Q=27$ ($\nu=1/3$), where states 
with 0, 1, or 2 CF excitons are considered.  
$E$ is the energy per particle, measured 
in units of $e^2/\epsilon l$, where $\epsilon$ is the dielectric constant of the host 
material and $l$ is the magnetic length.  The statistical uncertainty in the energy, 
coming from Monte Carlo sampling, is much smaller than the sizes 
of the dots.  The dashes are the exact eigenenergies obtained by numerical 
diagonalization.  The lowest energies in each $L$ sector are given in Table I, 
labeled $E^{(2)}$.}
\label{fig1}
\end{figure}

To understand higher energy excitations we consider a mixture of 
states containing zero, one or two excitons of composite fermions.  
In this case, the basis at $\nu^*=1$ contains several states at each 
$L$, producing, in turn,  many states at each $L$ for $\nu=1/3$.  
Diagonalization in this basis produces the spectrum shown in Fig.~(\ref{fig1}).
With this expanded basis, the lowest energy in each $L$ sector 
predicted by the CF theory, given in Table I, is practically exact.

It is noted that finite thickness, disorder, and Landau level
mixing, neglected in this article, 
must be taken into account for an accurate quantitative comparison with experiment.
The excitation energies observed in Raman experiments are 
in qualitative and semi-quantitative  
agreement with the CF theory at $\nu=1/3$ and $2/5$, and nicely demonstrate the 
formation of Landau-like levels in the intermediate fillings\cite{Pinczuk}.

\begin{table}
\begin{center}
\begin{tabular}{|c|c|c|c|c|}\hline 
$ L$ & $E^{exact}$ & $E^{(1)}_{CF}$ & $E^{(2)}_{CF}$ & error (\%)       \\ \hline
0 & -0.41062897 & -0.41039(2) & -0.41062(3) & -  \\ \hline
1 & -0.39609058 & - & -0.3957 (1) & 0.1 \\ \hline
2 & -0.39836847 & -0.3972 (1) & -0.3981 (1) & 0.07 \\ \hline
3 & -0.39985458 & -0.39916 (8) & -0.3996 (1) & 0.06 \\ \hline
4 & -0.40203892 & -0.4017 (1) & -0.40194 (5) & 0.02  \\ \hline
5 & -0.40354097 & -0.40324 (8) &  -0.40343 (4) & 0.03 \\ \hline
6 & -0.40268431 & -0.4023 (1) & -0.40261 (5) & 0.02 \\ \hline
7 & -0.4012351  & -0.40094 (8) & -0.4013 (2) & - \\ \hline
8 & -0.40115314 & -0.40081 (6) & -0.40115 (5) & - \\ \hline
9 & -0.40157815 & -0.40127 (5) & -0.4015 (1)  & - \\ \hline
10 & -0.40120351 & -0.40098 (8) & -0.4013 (1) & - \\ \hline
\end{tabular}
\caption{The second column gives the exact energy per particle for the
lowest energy at orbital angular momenta $L=0, 1, \cdots 10$ for
10 particles at $\nu=1/3$.  $E^{(1)}_{CF}$ is the energy of the state with
a single particle hole pair of composite fermion, obtained with the help
of a wave function with no adjustable parameters.  $E^{(2)}_{CF}$ is the
energy from Fig.~(\ref{fig1}), which is obtained by diagonalization in the
Fock space of all states containing 0, 1, or 2 pairs of particle hole
excitations of composite fermions.  The last column gives the \% error
for cases where $E^{(2)}_{CF}$ differs from $E^{exact}$ significantly.
All energies are quoted in units of $e^2/\epsilon l$.}
\label{tab2}
\end{center}
\end{table}

\section{Nature of the single quasiparticle of $\nu=1/3$}

\begin{figure}
\centerline{\psfig{figure=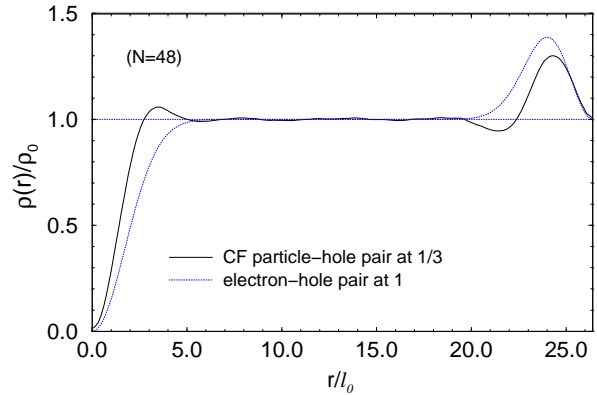,width=3.5in,angle=-90}}
\vspace{-0.5cm}
\caption{The density profile of a CF-particle at the South pole and a CF-hole at the North
pole at $\nu=1/3$ (solid line). For comparison, the density profile for an electron in the 
second LL (South pole) and a hole in the lowest LL (North
pole) at $\nu=1$ is also given (dotted line).  The distance is measured in units of the 
magnetic length, from the North pole to the South pole.  Source: 
Ref.~\protect\onlinecite{kwonthesis}.}
\label{fig2}
\end{figure}

\begin{figure}
\centerline{\psfig{figure=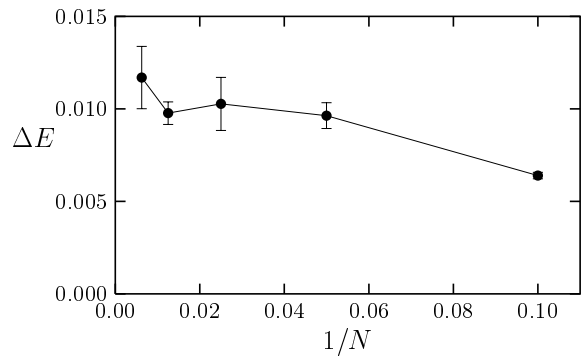,width=3.5in,angle=0}}
\vspace{-0.5cm}
\caption{$\Delta E$ is the energy difference between two wave functions 
for the $\nu=1/3$ quasiparticle (Laughlin's \cite{Laughlin}, and that 
based on composite fermions\cite{Jain,Jain2})  
as a function of $N$, the total number of 
particles.  The energy difference in the thermodynamic 
limit is $0.012(2) e^2/\epsilon l$, which is approximately 15\% of 
the energy of the quasiparticle ($\sim 0.07 e^2/\epsilon l$).}
\label{fig3}
\end{figure}

The following two facts about the quasiparticle at $\nu=1/3$ are puzzling at first
sight: (i) It has higher energy than the quasihole.  (ii) Its density profile is not 
the mirror image of the density profile of the quasihole, but has a 
strange smoke ring shape, with a minimum at the origin. 
They have found a natural explanation in the CF theory.
The asymmetry between the quasiparticle and the quasihole is ``obvious" in 
the CF theory, as the former is akin to an electron in the second LL,  
while the latter to a hole in the lowest LL.  That also explains why the quasiparticle 
has higher energy than the quasihole.  The density profile of the 
quasiparticle is qualitatively similar to density profile  
of an electron in the second LL, as shown in Fig.~(\ref{fig2})
(taken from \onlinecite{kwonthesis}).  
In addition to these remarkable qualitative insights, several microscopic  
calculations\cite{Kasner,Girlich,Melik} have demonstrated that the wave 
function for the quasiparticle motivated by 
the CF physics\cite{Jain,Jain2} has significantly lower energy 
than Laughlin's\cite{Laughlin}.  Fig.~(\ref{fig3}) 
shows the energy difference between the two wave functions 
for up to $N=160$ particles\cite{Jeon} in the disk 
geometry; the former has approximately 15\% lower energy.
The quantitative discrepancy between the two
approaches grows with the number of quasiparticles \cite{Jeon}.

These results establish that {\em the ``quasiparticle" of 1/3 is in fact a 
composite fermion in the second level}.  That already provides a strong hint 
that the state in the filling factor range $2/5 > \nu > 1/3$
is to be interpreted in terms of composite fermions partially 
occupying the second level.

\section{$2/5>\nu>1/3$}

The CF theory has been tested in the past at $\nu=n/(2pn+1)$, as
well as at fractions in between.   For the region of our interest,
where $2>\nu^*>1$, the problem maps into 
$\bar N=N-(2Q^*+1)$ composite fermions in $L^*=Q^*+1$ shell 
(i.e., the second level), with $Q^*=Q-N+1$.

\begin{figure}
\centerline{\psfig{figure=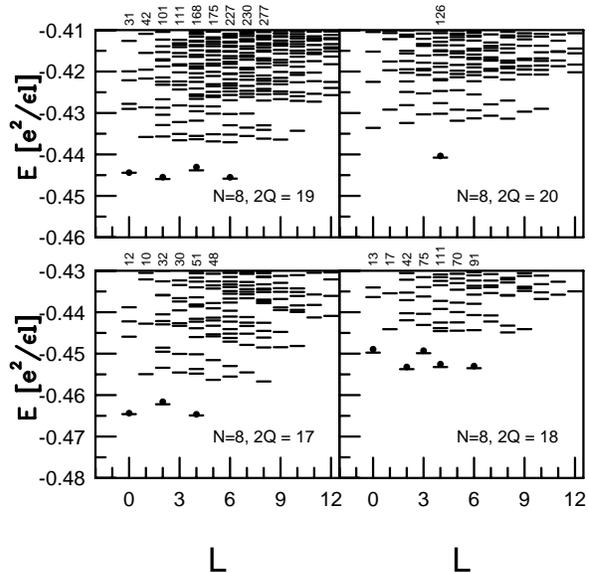,width=3.5in,angle=0}}
\vspace{-0.5cm}
\caption{Comparison between the CF theory (dots) and exact spectrum (dashes) for 
$N=8$ particles in the filling factor range $2/5>\nu>1/3$, with the flux through 
the surface of the sphere given by $2Q hc/e$.
$E$ is the energy per particle, in units of $e^2/\epsilon l$.
$L$ is the total orbital angular momentum, with the 
total number of multiplets at several values of $L$ given at the top axis.
Source: Ref.~\protect\onlinecite{Jain3}.}
\label{fig4}
\end{figure}

Fig.~(\ref{fig4}) shows a study in the filling factor range $2/5>\nu>1/3$ 
for $N=8$ particles\cite{Jain3}.  As the flux changes from 
$2Q=20$ to $2Q=17$, the number of composite fermions in the 
second level, $\bar N$, changes from $\bar N=1$ to $\bar N=4$.  
Fig.~(\ref{fig4}) presents results for  
$N=12$ particles at flux $2Q=29$, which 
map into $N=12$ composite fermions at $2Q^*=7$.  The 
lowest level can accommodate 8 composite fermions and the second level 
has $\bar N=4$ composite fermions.  In all of these figures, the 
dashes are the exact energies\cite{Rezayi} and the dots are the energies predicted by 
the CF theory.   The CF energies in Fig.~(\ref{fig4}) 
are essentially the same as those in Fig.~3 of 
Ref.~\onlinecite{Mandal}, but calculated with greater accuracy. 
(Slightly lower energies than those quoted in Ref.~\onlinecite{Mandal}
are obtained by varying the function used for importance sampling in our Monte Carlo.)
The following points are noteworthy.

The CF theory makes the following prediction for the total angular 
momentum quantum numbers for the low energy states:

$\bullet$ $(N,Q)=(8,10)$ maps into $\bar N=1$ at $L^*=4$.
Here we have only one multiplet with total angular momentum $L=4$.

$\bullet$ $(N,Q)=(8,\frac{19}{2})$ maps into $\bar N=2$ at $L^*=\frac{7}{2}$.
The possible total angular momenta, within the constraints of the 
Pauli principle, are given by
$$\frac{7}{2}\otimes \frac{7}{2}=0\oplus 2 \oplus 4 \oplus 6 $$

$\bullet$ $(N,Q)=(8,9)$ maps into $\bar N=3$ at $L^*=3$.
The possible total angular momenta, incorporating the Pauli principle, are
$$3 \otimes 3 \otimes 3 =0\oplus 2 \oplus 3 \oplus 4 \oplus 6$$

$\bullet$ $(N,Q)=(8,\frac{17}{2})$ maps into $\bar N=4$ at $L^*=\frac{5}{2}$.
The possible total angular momenta now are
$$\frac{5}{2}\otimes \frac{5}{2} \otimes \frac{5}{2} \otimes \frac{5}{2} =
0\oplus 2 \oplus 4$$
In this case, the degeneracy of the angular momentum shell is $2L^*+1=6$, 
so it would have been easier to consider two fermion holes in the $L^*=\frac{5}{2}$
shell.

$\bullet$ $(N,Q)=(12,\frac{29}{2})$ is equivalent to  
$\bar N=4$ fermions in the 9/2 angular momentum shell.  Here we have 
\begin{eqnarray}
\frac{9}{2}\otimes && \frac{9}{2}\otimes \frac{9}{2}\otimes \frac{9}{2} =
\nonumber \\
&& 0^2\oplus  2^2 \oplus 3 \oplus 4^3 \oplus 5 \oplus 6^3 \oplus 7 \oplus 
8^2 \oplus 9 \oplus 10 \oplus 12 
\nonumber
\end{eqnarray}

\begin{figure}
\centerline{\psfig{figure=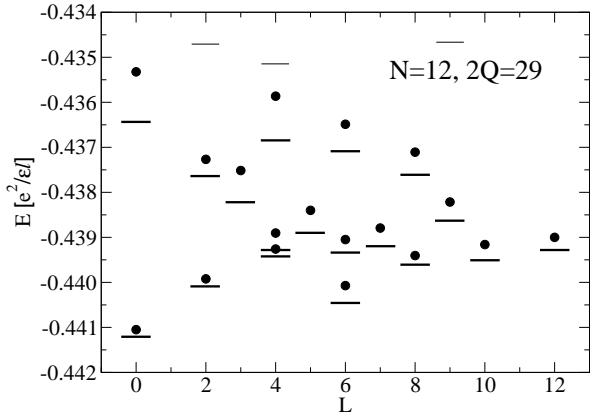,width=3.5in,angle=-90}}
\vspace{-0.5cm}
\caption{The dots show the spectrum of low energy states predicted by the 
CF theory for $N=12$ electrons at flux $2Q=29$.  $E$ is the energy per particle, measured 
in units of $e^2/\epsilon l$, where $\epsilon$ is the dielectric constant of the host 
material and $l$ is the magnetic length.  The statistical uncertainty in the energy, 
coming from Monte Carlo sampling, is roughly equal to the size of the 
dots.  The dashes are the exact eigenenergies obtained by numerical diagonalization.
The darker dashes form a low energy band of states, which have a one to one 
correspondence with the states predicted by the CF theory.
The exact eigenenergies are taken from Ref.~\protect\onlinecite{Rezayi}.}
\label{fig5}
\end{figure}

In all cases, the angular momentum quantum numbers predicted by 
the CF theory match perfectly the quantum numbers of the states forming 
the low energy bands in the exact spectra of Figs.~(\ref{fig4}) and (\ref{fig5}).  
These results give a clear evidence that the strongly 
interacting system of electrons in the lowest Landau level resembles 
a system of weakly interacting fermions at an effective magnetic field.

The most convincing theoretical verification of the CF theory comes from 
comparing the exact energies (dashes in Figs.~\ref{fig4} and \ref{fig5})
with those predicted by the CF theory (dots).  For example, for the 
12 particle system the lowest energy in each angular momentum sector 
is predicted with an accuracy of $\sim 0.1$\% or better (Table II).  
(For higher energies the accuracy is somewhat worse, but still very good.  The reason 
for the discrepancy is the neglect of mixing between different CF levels.)

The interpretation of the actual state in the range $2/5 > \nu > 1/3$ in 
terms of composite fermions (Fig.~\ref{fig0}) is thus established.
A crucial aspect of composite fermions is that they are {\em weakly 
interacting}, which is what makes them the ``true" quasiparticles of 
the FQHE, in the same sense as the Landau quasiparticles are the true 
quasiparticles for the normal Fermi 
liquid or the Cooper pairs for the superconductor.  The weakness of the 
residual interaction can be seen in the above studies.   If 
the composite fermions were non-interacting then all dots would be degenerate.
The splitting between various states is therefore a measure of the residual interaction 
between composite fermions.  Take, for example, the 12 particle system at $\nu=4/11$, 
the low energy band (dark dashes) of which is analogous to that of four 
composite fermions at $\nu=1/3$.  The width of the band in Fig.~(\ref{fig4}) 
is $\sim 0.07 e^2/\ell$ (note that the energies shown are {\em single} particle
energies), which, when compared to  
$\sim 0.7 e^2/\ell$, the width of the band for 4 electrons at 1/3, 
indicates that the residual interaction between the composite fermions is an 
order of magnitude weaker than the interaction between electrons at $\nu=1/3$.

\begin{table}
\begin{center}
\begin{tabular}{|l|l|l|} \hline
$ L$ & $E^{exact}$ & $E_{CF}$       \\ \hline
0 &  -0.441214 & -0.441051(89) \\ \cline{2-3}
  &  -0.436440 & -0.435323(69) \\ \hline
2 &  -0.440457 & -0.439922(86) \\ \cline{2-3}
  &  -0.437646 & -0.437268(59) \\ \hline
3 &  -0.438226 & -0.437516(72) \\ \hline
4 &  -0.439422 & -0.439260(46) \\ \cline{2-3}
  &  -0.439280 & -0.438904(58) \\ \cline{2-3}
  &  -0.436844 & -0.435864(49) \\ \hline
5 &  -0.438904 & -0.438400(63) \\ \hline
6 &  -0.440547 & -0.440072(75) \\ \cline{2-3}
  &  -0.439337 & -0.439050(42) \\ \cline{2-3}
  &  -0.437093 & -0.436488(80) \\ \hline
7 &  -0.439190 & -0.438794(79) \\ \hline
8 &  -0.439613 & -0.439404(98) \\ \cline{2-3}
  &  -0.437615 & -0.437108(88) \\ \hline
9 &  -0.438632 & -0.438215(10) \\ \hline
10 & -0.439507 & -0.439160(50) \\ \hline
12 & -0.439287 & -0.439000(11) \\ \hline
\end{tabular}
\caption{Comparison of the CF prediction for the
lowest energy band at $\nu=4/11$ for
$N=12$ particles with the exact eigenenergies from Ref.~\protect\onlinecite{Rezayi}.
The number of multiplets at each $L$ is predicted correctly by
the CF theory.}
\label{tab1}
\end{center}
\end{table}

\section{$\nu=4/11$}

If the composite fermions were completely non-interacting, then 
there would be no FQHE at $\nu^*=1+1/3$ ($\nu=4/11$), 
just as there would be no FQHE for electrons 
at $\nu=1+1/3$ if the Coulomb interaction were switched off.  
However, there is a weak residual interaction between composite fermions,
which may possibly cause a gap to open, thereby producing a FQHE at $\nu=4/11$.  
This provides a simple scenario for FQHE at 4/11 (and a host of other 
new fractions).  However, going beyond mere scenarios,  
the question is whether the residual interaction between 
composite fermions is sufficiently strongly repulsive at short 
distances to produce a FQHE of composite fermions here.  
A quantitative investigation of the issue requires a determination of   
minute energy differences accurately.  Given that we are dealing with a 
strongly correlated many body state, the situation might seem intractable 
at first, 
but, as seen in Fig.~(\ref{fig4}) and Table II,  the wave functions of the 
CF theory possess the desired accuracy.

The systems at flux $2Q=(11/4)N-4$ are believed to represent the thermodynamic 
filling $\nu=4/11$.  Thus, the 12 particle system at $2Q=29$ and the 
8 particle system at $2Q=18$ are relevant for possible FQHE at $\nu=4/11$.
(Note that the quantum numbers of the low energy states in exact diagonalization 
are identical to the quantum numbers of all states of $\bar N$ fermions at 
$\nu=1/3$.) For $N=12$, the ground state appears to be incompressible, 
with an $L=0$ ground state and a gap to excitations.  
However, incompressibility in the thermodynamic limit requires 
one to study how the gap behaves as a function of $N$ and 
to show that it survives in the $N^{-1}\rightarrow 0$ limit.
That is particularly crucial here because, as W\'{o}js and Quinn\cite{Quinn} 
have noted, the exact spectrum for $N=8$ particles at $\nu=4/11$ 
(see Fig.~\ref{fig4}) indicates a {\em compressible} ground state.
Also, while 12 may appear to be a reasonably large number 
of particles, the relevant number is $\bar N=4$, the number of composite 
fermions in the second level, just as at filling factor $\nu=1+1/3$ the 
relevant number is the number of electrons in the second LL, with the 
electrons filling up the lowest LL being inert.  To further investigate 
if the incompressibility persists in larger systems,
Mandal and Jain\cite{Mandal} carried out CF diagonalization 
for $N=16$, 20, and 24 particles, for which $\bar N=5$, 6, and 7.  
It was found that as $N$ changes in steps of four, the system alternates between 
compressible ($N=8$, 16, 24) and incompressible ($N=12$, 20).
The fact that two of the systems suggest an incompressible state makes the 
interpretation of the results somewhat less certain, 
but we do not know of any justification for disregarding the other systems.
On the basis of the facts that (i) 
for all securely understood FQHE states to date, the spectrum reflects an 
incompressible state at {\em all} allowed values of $N$; 
and  (ii) for states that are compressible in the thermodynamic limit, 
the spectrum can at times resemble an incompressible 
state for finite systems\cite{Fano}, it was concluded\cite{Mandal} 
that the available results do not support incompressibility at $\nu=4/11$ for 
the model considered.

This conclusion is at odds with the experimental 
observation\cite{Pan} of FQHE at $\nu=4/11$. At the same time, the comparisons 
with exact results show that the CF theory is very accurate.  
How does one reconcile these two facts?   Several possibilities come to mind. (i)  
A partially spin polarized FQHE state at $\nu=4/11$ has been found
to be stable theoretically\cite{Park}, and one may wonder if the experimental 
4/11 state could be partially polarized.  From our calculations, it is 
possible to obtain a crude estimate of the crossover magnetic field. 
The Coulomb energy of the partially polarized state is lower by approximately 
$0.005 e^2/\epsilon l$ per particle compared to the Coulomb energy of the 
fully polarized state.  Which one is the ground state depends on the Zeeman 
splitting: at sufficiently low Zeeman energies the 
partially polarized state is the ground state, but as the Zeeman energy 
is increased, at some point there will be a transition into  
the fully polarized state.  Because 
the partially polarized state has one-fourth of the electron spins flipped, 
the crossover magnetic field is given by the formula $0.005 e^2/\epsilon l=E_Z/4$,
where $E_Z$ is the energy required to flip a single spin.  For GaAs parameters, the 
crossover magnetic field is estimated to be $\approx$ 12 T for the idealized model,
which is fairly high.  However, the $\nu=4/11$ FQHE has been observed at higher 
magnetic fields, which appears to rule out a partially polarized state here.
(ii) Another possibility is that the assumption of neglecting finite thickness 
or Landau level mixing is not sufficiently accurate for the problem of the 
4/11 FQHE.  These effects modify the form of the
interaction between the electrons.  Usually, a slight modification
of the interaction only causes a {\em quantitative} correction, but
because of the extremely tiny energy scales governing the physics here,
it is possible that it may alter the ordering
of the energy levels to produce incompressibility.  Unfortunately, the neglected effects
are not understood at the same level of accuracy as the CF theory of the FQHE.
(iii)  We have explored above a 4/11 FQHE state that is analogous
to $\nu^*=1+1/3$ FQHE of composite fermions.  
It is, in principle, possible that the 4/11 FQHE state is an entirely new type of 
incompressible state of composite fermions.

While we must await the resolution of the 4/11 enigma, our calculations make 
a compelling case that the CF theory provides the correct framework
for an investigation of the issue.

Partial support of this research by the National Science Foundation under grants
no. DGE-9987589 (IGERT) and DMR-0240458 is acknowledged.  We are grateful to 
E.H. Rezayi for sharing with us his exact diagonalization results.

\end{document}